\documentclass[prb,aps,preprint,showpacs]{revtex4}

\usepackage{amsmath,amssymb,bm}
\usepackage{graphicx}
\usepackage{caption}
\usepackage{mathrsfs}
\usepackage{subcaption}
\usepackage[colorlinks=true, linkcolor=blue, citecolor=blue, urlcolor=blue]{hyperref}
\hypersetup{hypertex=true,
colorlinks=true,
linkcolor=blue,
anchorcolor=blue,
citecolor=blue}
\usepackage{color}
\usepackage{newtxmath}
\usepackage{newtxtext}
\renewcommand\d{\partial}

\newcommand\q{{\boldsymbol{q}}}
\renewcommand\k{{\boldsymbol{k}} }

\newcommand\<{\langle}
\renewcommand\>{\rangle}

\newcommand\Tr{\textrm{Tr}}

\begin{document}
\title{Phase transition of Kitaev spin liquid described by quantum geometric tensor}
\author{Meng-Meng Lu}
\author{Zheng-Chuan Wang}
\email{wangzc@ucas.ac.cn}

\affiliation{School of Physical Sciences, University of Chinese Academy of Sciences, Beijing 100049, China}

\begin{abstract}
   We investigate the topological phase transition of Kitaev spin liquid in an external magnetic field by calculating the Berry curvature and the Fubini-Study metric. 
   Employing Jordan-Wigner transformation and effective perturbative theory to transform the Hamiltonian into fermionic quadratic form, the Berry curvature is calculated by choosing the effective magnetic field as the parameter, and we find that the xy-component of the Berry curvature has the same behavior around the critical lines with the phase diagram and the behavior of Berry curvature around the critical line will not be influenced by local perturbation, i.e. it has the robustness against the local perturbation.
   Especially, we relate the Berry curvature with the derivative of effective magnetic susceptibility which can be regarded as the signature of topological phase transition besides, we related the second nonlinear susceptibility with the non-Abelian Berry connection.
   Then we analytically calculate the generalized Berry curvature in the mixed state called mean Uhlmann curvature which can be related with the spectral function, the curves that mean Uhlmann curvature changing temperature with different coupling constant reveal that it will have an extrema when adjust $J_x$ from $A$ phase to $B$ phase.
   At last we analytically calculate the quantum geometric tensor in the effective magnetic field space whose imaginary part is the Berry curvature and real part is the Fubini-Study metric, and we find that the zz-component of Fubini-Study metric and the phase diagram are highly correlated which will peak at the cross point of three phases, then the Fubini-Study metric is extended to the finite temperature with arising an additional term called Fisher-Rao metric caused by mixed state.

 \end{abstract}

\maketitle  

\section*{I. Introduction}
\label{Sec.I}
The quantum spin liquid is an exotic magnetic system without any long-range order even near zero temperature which is extremely disorganized like the paramagnetic system. But the difference from paramagnetic system and also the key property is that the QSL state has long-range quantum entanglement, 
which were first proposed by Adnerson in 1973\cite{anderson1973}, the so-called RVB state, when the hight-temperature superconductivity was discovered in experiment\cite{anderson1987}, the RVB state is able to be obtained by Gutzwiller projection for BCS function. 
The QSL has two important properties: long-rangle entanglement and fractional excitations, which is beyond the traditional condensed matter physical frame without any symmetry breaking\cite{Yi Zhou,Balents2010}.
In other words, it can't be described by the theory of spontaneous symmetry breaking proposed by Landau because it doesn't have the local order parameter.

As the QSL concept was proposed for now, there are many novel theories applied in this field trying to describe it, such as the effective field theory\cite{Yi Zhou} and the topological order\cite{XGW1989,XGW1991,XGW2002}.
After the introduction of fractional quantum Hall effect(FQHE), the topological matter and the topological phase transition have aroused great interest\cite{Laughlin1983} which can be characterized by fractional excitation and the ground state degeneracy intead of the symmetry breaking. After then the topological field theory was proposed to characterized the FQHE with the Chern-Simons action briefly.
Based on the above theories X.G. Wen introduced a generalized order called the topological order related to the degree of ground state degeneracy to characterize the topological phase transition giving an exact description about the chiral spin liquid\cite{CSL1989,CSL1987}. In order to develop the generalized theory of QSL,with the assumption of mean field theory, Wen proposed the projective symmetry group and SU(2) gauge field theory which classifies the QSL by symmetry\cite{XGW2002}.
 
Among QSL models, the toric code model introdunced by Kitaev\cite{kitaev2003} is an exactly solvable spin system located on a 2-$D$ square lattice with peroidic boundary condition which was designed to realize the quantum-correcting code.
After introducing by Kitaev, it was found that it has the non-Abelian anyon excitation whose ground state is the $\mathbb{Z}_2$ spin liquid state with 4 degree of topological degeneracy.
Since the model was proposed, a lot of works have been done to study it and its extended dynamical quantum phase transitions\cite{toric2008,toric2009,toric2019,toric2023}. In 2021, Google Quantum AI group realized the toric code model by superconductivity quantum circuit including 31 spins\cite{google2021}. They verified that the entanglement entropy follows the area law whose measured value is closed to the theoretical value $-\ln2$.
At the same time, Lukin group from Harvard university realized the toric code model on the Kagome lattice with 219 Rydberg atoms trapped in the light lattice\cite{harvard2021}. These above experiments show that the quantum simulator makes the QSL realization possible.\par
In 2006 Kitaev introduced another exactly solvable honeycomb model\cite{kitaev2006} called Kitaev honeycomb model(KHM). The KHM is a spin system with strongly anisotropic interaction which looks like the Heisenberg interaction with the difference that the KHM's interaction is bond-dependent.
The nontrivial ground state of KHM is the QSL state with non-Abelian anyon excitation and it has rich phase diagram and many of its physical quantities can be calculated exactly. When applying an uniform magnetic field on this model, the phase diagram will show both Abelian and non-Abelian gapped phases\cite{kitaev2023}.In the non-Abelian phase, it wll emerge the gapped Majorana bound state corresponding the non-Abelian anyon\cite{kitaev2006,kitaev2023,Lahtinen2008}.
Besides, the external field will break the time-reversal symmetry of the KHM which can be solved approximatly by using perturbative theory assuming the external field is small enough\cite{kitaev2006}.
Then we can find that the Chern number of KHM is non-vanished hence one can claim it realizes the non-Abelian Ising topological order with a chiral bound state\cite{kitaev2006}.
   
Kitaev spin liquid is highly correlated with the materials in the experiment which opens a door to find the QSL state in marerials. In experiments, there are many candinate materials for Kitaev QSL, \textbf{Na$_2$IrO$_3$}\cite{Na2IrO3}, \textbf{Li$_2$IrO$_3$}\cite{Li2IrO3}, and the most attracting $\alpha$-\textbf{RuCl$_3$}\cite{RuCl3}, all which have the strongly anisotropic interaction but show the weak magnetic order at low temperature. 
The pure Ktaev-type interaction is diffcult to realize because all the candinate materials' Kitaev interaction is not strong enough, i.e. there are other kinds of interactions such as Heisenberg interaction, $\Gamma$ interaction.

On theory, the phase transition in Kitaev QSL from one phase to another is the topological phase transition beyond Landau paradigm. Recently, much effort has been done to analyse the phase transition about the KHM in the vertox-free sector\cite{iop2019} according to the Berry curvature(BC) of the critical KHM.
Then they extended this work to the finite-temperature of KHM with the Uhlmann curvature and number including the temperature, resulted that the Uhlmann curvature has a peak at the critical point of transition and the mean Uhlmann curvature describes a cross-over with temperature rising\cite{uhlmann2019}.
However, physically, the curvature about geometrical property is not a directly measurable quantity without physical quantity corresponding to. Hence, we proposed a plan to deal with this trouble in which we relate the BC and the physical quantity by the effective magnetic susceptibility and found the explicit relation between the magnetic susceptibility and the Berry connection by linear response theory.
Then we calculated the Fubini-Study metric which relates to the construct factor and forms the quantum geometric tensor together with BC, found that the Fubini-Study metric is a signature of phase transition.

The paper is organized as follows:
   In Sec.II we derived the BC in the parameter space of Kitaev model and reduced the Hamiltonian into a closed form organized as the summation of spin dot the effective magnetic field. We obtained the relation between the effective spin susceptibility in the effective magnetic field and the BC in the $\mathbf{B}$ manifold by Lehmann representation. Furthermore, we derived the second nonlinear susceptibility by Lehmann representation and established the relation with Berry connection. 
   In Sec.III we generalized the BC to the mixed state as \cite{iop2019,uhlmann2019,invariantsystem,dissipative} did which is related with the  spectral function and the effective magnetic susceptibility. Then we numerically calculated the mean Uhlmann curvature.
   In Sec.IV we exactly calculated the Fubini-Study metric which organizes the quantum geometric tensor with Berry curvature and analytically obtained that the each components of geometric tensor and Fubini-Study metric changing with $J_x,J_y$. Then we calculated the Bures metric which is the generalization from pure states to  mixed states of Fubini-Study metric. Then we simplified the second term of Bures metric into a compact form which is similar to mean Uhlmann curvature.

   \section*{II. Susceptibility and Berry curvature}
   \label{Sec.II}
   \subsection*{A. Spin correlation and effective magnetic susceptibility}
\begin{figure}[!htb]
   \setlength{\abovecaptionskip}{-0.6cm}
\setlength{\belowcaptionskip}{0.5cm}
   \includegraphics[scale=0.25]{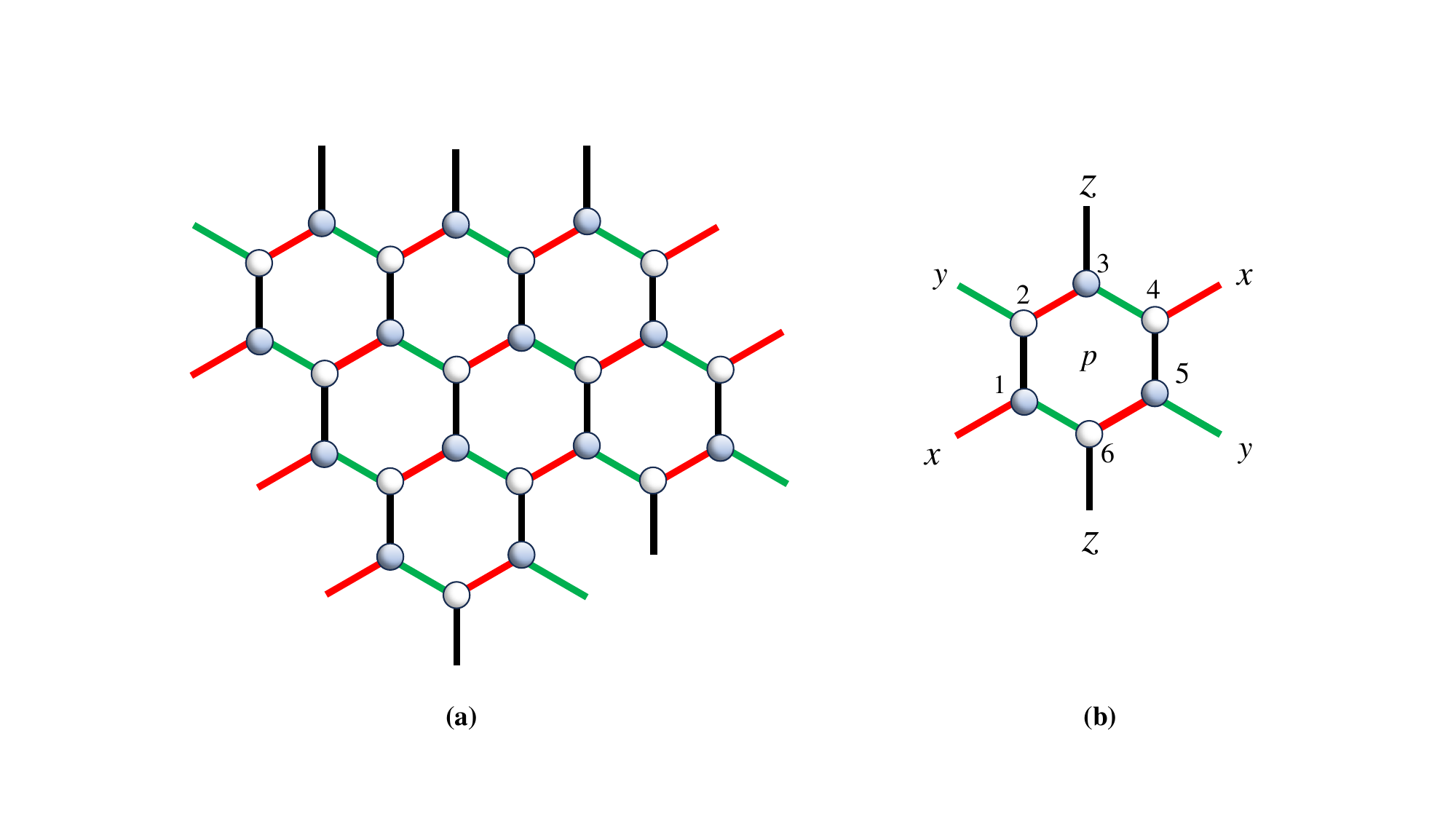}
   \centering
   \caption{Honeycomb model. In Fig.(b), we can see that the red, the green and black line are the x-link,y-link,and the z-link, respectively.}
   \label{Honeycomb}
\end{figure}

The system we study is the Kitaev spin liquid model\cite{kitaev2006} which is a spin-$\frac12$ system located on honeycomb lattice whose interaction is bond-dependent and the Hamiltonian is :
\begin{equation}
   H_0=-\sum_{\alpha\in{x,y,z}}\sum_{i,j}J_\alpha\hat{\sigma}^\alpha_i\hat{\sigma}^\alpha_j,
\end{equation}
where $J_\alpha$ is the anisotropic interaction strength and the pairing form of spin is bond-dependent as Fig.\ref{Honeycomb}.

The pure Kitaev spin liquid\cite{kitaev2006} has been studied sufficiently by many others, here we are paying attention to the Kitaev spin liquid in an external magnetic field. The Hamiltonian can be written as:
\begin{equation}
   H=H_0+H',
\end{equation}
with a Zeeman perturbation term $H'$
\begin{equation}
   H'=-\sum_{i}h_x\hat\sigma_i^x+h_y\hat\sigma_i^y+h_z\hat\sigma_i^z.
\end{equation}
The perturbation term is equivalent to a three-body interaction\cite{JWJPA2008,iop2019,uhlmann2019} in terms of:
\begin{equation}
   H'=-K\sum_p\sum_{\ell=1}^4P_p^{(\ell)},
\end{equation}
with
\begin{equation}
   \sum_{\ell=1}^4P_p^{(\ell)}=\sigma_1^x\sigma_6^y\sigma_5^z+\sigma_2^z\sigma_3^y\sigma_4^x+\sigma_1^y\sigma_2^x\sigma_3^z+\sigma_4^y\sigma_5^x\sigma_6^z.
\end{equation}
  By means of Jordan-Wigner transformation\cite{JWJPA2008,JWPRB2007,JWPRL2007} the Hamiltonian of Kitaev model can be written as:
  \begin{equation}
   H=\frac{1}{2}\sum_{{\k} }\left(c_{{\k} }^{\dagger},c_{-{\k} }\right)H_{{\k} }\left(\begin{matrix}c_{{\k} }\\c_{-{\k} }^{\dagger}\end{matrix}\right)=\frac{1}{2}\sum_{k}\hat{\psi}_{{\k} }^{\dagger}H_{{\k} }\hat{\psi}_{{\k} },
  \end{equation}
where $\psi_\k,\psi_\k^\dagger $ are the Nambu spinors and the Hamiltonian in the $\k$-space is given by:

   \begin{equation}
      H_\k=\begin{pmatrix}
         \xi_\k &\Delta_\k \\
         \Delta^*_\k &-\xi_\k 
      \end{pmatrix}=\hat\sigma_\cdot\mathbf{B}_\k,\label{effectivemagnetic}
   \end{equation}
   with
   \begin{equation}
      \mathbf{B}_\k=\left(\alpha_\k ,-\beta_\k ,\xi_\k  \right)\label{B_k},~~\Delta_\k=\alpha_\k+i\beta_\k,
   \end{equation}
   and\cite{uhlmann2019}
   \begin{equation}
      \begin{split}
         \xi_\k&=2J_x\cos k_x+2J_y\cos k_y+2J_z\\
         \beta_\k&=2(J_x\sin k_x+J_y\sin k_y)\\
         \alpha_\k&=4K(\sin k_x-\sin k_y),\label{factors}
      \end{split}
   \end{equation}
where $\mathbf{B}_\k$ we defined is the effective magnetic field and $\vec\sigma$ is Pauli matrix. Hence, it's easy to calculate its two energy spectrums as:
\begin{equation}
   E_{\k,0}=-\varepsilon_\k,~~~E_{\k,1}=\varepsilon_\k,
\end{equation}
with:
\begin{equation}
   \varepsilon_\k=\sqrt{\alpha_\k^2+\beta_\k^2+\xi_\k^2}.
\end{equation}
 In the form of second quantization, the total Hamiltonian is given by
\begin{equation}
   H=\sum_\k \left(\hat\psi_\k ^\dagger\vec\sigma\hat\psi_\k \right)\cdot\mathbf{B}_\k=\sum_\k \vec{\sigma}_\k \cdot\mathbf{B}_\k,
\end{equation}
where $\vec\sigma_\k =\hat\psi_\k ^\dagger\hat\sigma\hat\psi_\k $ is the spin density operator in the $\k $ space. In the $\{\mathbf{B}_\k\}$ parameter space, we can write the spectral BC\cite{uhlmann2019} via definition
\begin{equation}
   \begin{split}
      \mathcal{F}_{\k ,\alpha\beta}&=i\big[
     \< \partial_{\k \alpha}\psi_0|\partial_{\k\beta}\psi_0\>-\<\partial_{\k\beta}\psi_0|\partial_{\k \alpha}\psi_0\>
   \big]\\
   &=-2\mathbf{Im}~\left[
      \<\partial_{\k \alpha}\psi_0|\partial_{\k\beta}\psi_0\>
   \right]\label{curvature}
   \end{split}
\end{equation}
in which for the sake of simplicity, we use $ \partial_{\k \alpha}$ to denote $\partial/\partial B_{{\k \alpha}}$, use $\mathbf{Im}$ to extract the imaginary part and $|\psi_0\>$ is the ground state and the total Berry curvature should be written as the summation of the spectral BC(we placed the derivation of the total Berry curvature  in the Appendix C.):
\begin{equation}
   \mathcal{F}_{\alpha\beta}=\sum_{\k}\mathcal{F}_{\k,\alpha\beta}.
\end{equation}
Then we represent the n-th eigenstate with $|\psi_n\>$ and based on following identity:
\begin{equation}
   \<\psi_0|\partial_{\k ,\alpha}|\psi_n\>=\frac{\<\psi_0|\partial_{\k ,\alpha}H|\psi_n\>}{E_n-E_0}\label{identity},
\end{equation}
insert a set of complete orthonormal basis into Eq.(\ref{curvature}) and we'll obtain the relation between curvature Eq.(\ref{curvature}) and Eq.(\ref{identity}):

\begin{equation}
   \begin{split}
      \mathcal{F}_{\k ,\alpha\beta}&=-2\mathbf{Im}~\frac{\<0_\k|\partial_{\k \alpha}H_\k|1_\k\>\<1_\k|\partial_{\k\beta}H_\k|0_\k\>}{(E_{\k,0}-E_{\k,1})^2}\\
      &=-2\mathbf{Im}~\frac{\<0_\k|\sigma_{\k \alpha}|1_\k\>\<1_\k|\sigma_{\k\beta}|0_\k\>}{(E_{\k,0}-E_{\k,1})^2},
   \end{split}
\end{equation}
where $|0_\k\>$ is the ground state of $H_\k$ and $|1_\k\>$ is the excited state given in Eq.(\ref{ground state}) and Eq.(\ref{excited state}).
  Note that the spin density correlation in the $\k$-space is given by:
\begin{equation}
   \chi_{\k ,\alpha\beta}=-i\Theta(t)\<[\sigma_{\k \alpha}(t),\sigma_{\k\beta}(0)]\>_0.\label{definition of susceptibility}
\end{equation}
We can note that the Jordan-Wigner transformation aims to transform the spin into the quasifermion so, it means the correlation of quasifermion spins.Then we expand the commutator and insert a set of complete orthonormal basis under the Heisenberg picture and we'll obtain the Lehmann representation of the susceptibility by doing Fourier transformation\cite{spinberry}
\begin{equation}
   \begin{split}
      \chi_{\k ,\alpha\beta}(\omega)&=\int\mathrm{d}t \mathbf{e}^{i\omega t}\chi_{\k \alpha,\k \beta}(t) \\
      &=\lim_{\eta\rightarrow0}\left(\frac{\<0_\k|\sigma_{\k \alpha}|1_\k\>\<1_\k|\sigma_{\k \beta}|0_\k\>}{\omega+i\eta-(E_{\k,1}-E_{\k,0})}\right.\left.-\frac{\<0_\k|\sigma_{\k \beta}|1_\k\>\<1_\k|\sigma_{\k \alpha}|0_\k\>}{\omega+i\eta-(E_{\k,0}-E_{\k,1})}\right).\label{FT of susceptibility}
   \end{split}
\end{equation}
The susceptibility and its derivative at $\omega=0$ are :
\begin{equation}
   \begin{split}
      \chi_{\k ,\alpha\beta}(0)&=-2\mathbf{Re}~\frac{\<0_\k|\sigma_{\k \alpha}|1_\k\>\<1_\k|\sigma_{\k \beta}|0_\k\>}{E_{\k,0}-E_{\k,1}}\\
      \frac{\partial}{\partial\omega}\chi_{\k ,\alpha\beta}(0)&=-2i\mathbf{Im}~\frac{\<0_\k|\sigma_{\k \alpha}|1_\k\>\<1_\k|\sigma_{\k \beta}|0_\k\>}{(E_{\k,0}-E_{\k,1})^2}.
   \end{split}
\end{equation}
Here, the effective spin magnetic susceptibility was related with the BC which is given by:
\begin{equation}
   \mathcal{F}_{\k ,\alpha\beta}=-i\frac{\partial}{\partial\omega}\chi_{\k ,\alpha\beta}(0)=-i\frac{\d}{\d\omega}\mathbf{Im}\chi_{\k,\alpha\beta}(0).\label{susceptibility}
\end{equation}
We can see that in limit of the low frequency, under the $\mathbf{B}_\k$ manifold the BC directly relates to the 1-order spin magnetic susceptibility of the quasifermion system.

\subsection*{B. Calculation of Berry curvature}
In this part, we'll show the general calculation method of BC by choosing the effective magnetic field space and for the sake of simplicity, we manipulate the following replacement:
\begin{equation}
   \frac{\partial}{\partial B_{\k,\alpha}}=\partial_{\k \alpha},\label{partial derivative}
\end{equation}
the Abelian Berry connection and non-Abelian Berry connection are given by:
\begin{equation}
   \mathcal{A}_{\k\alpha}=-i\<\psi_0|\partial_{\k\alpha}|\psi_0\>\quad \mathcal{A}_{\k \alpha}^{ij}=-i\<\psi_i|\partial_{\k \alpha}|\psi_j\>.\label{connection}
\end{equation}
For the Hamiltonian Eq.(\ref{effectivemagnetic}), it's easy to obtain the spectral curvature proposed by Bascone et al.\cite{uhlmann2019}
\begin{equation}
   \mathcal{F}_{\k,\alpha\beta}=\partial_{\k\alpha}\mathcal{A}_{\k\beta}-\partial_{\k\beta}\mathcal{A}_{\k\alpha}.
\end{equation}
This is a general method of calculating BC in many textbook. The form of BC is like the spin in an external mangetic field and the components of spectral BC 
\begin{equation}
   \begin{split}
             \mathcal{F}_{\k ,xy}&=\frac{B_{\k,z}}{2\varepsilon_\k^3}=\frac{\xi_{\k }}{(\alpha_\k ^2+\beta_\k ^2+\xi_\k ^2)^{3/2}}=-\mathcal{F}_{\k ,yx}\\
             \mathcal{F}_{\k ,yz}&=\frac{B_{\k,x}}{2\varepsilon_\k^3}=\frac{\alpha_{\k }}{(\alpha_\k ^2+\beta_\k ^2+\xi_\k ^2)^{3/2}}=-\mathcal{F}_{\k ,zy}\\
             \mathcal{F}_{\k ,zx}&=\frac{B_{\k,y}}{2\varepsilon_\k^3}=\frac{-\beta_{\k }}{(\alpha_\k ^2+\beta_\k ^2+\xi_\k ^2)^{3/2}}=-\mathcal{F}_{\k ,xz},
   \end{split}
\end{equation}
and the diagonalized elements of BC matrix are zero, i.e.:
\begin{equation}
   \mathcal{F}_{\k }=
  \left( \begin{array}{ccc}
      0&\mathcal{F}_{\k,xy}&-\mathcal{F}_{\k,zx}\\\\
      -\mathcal{F}_{\k,xy}&0&\mathcal{F}_{\k,yz}\\\\
      \mathcal{F}_{\k,zx}&-\mathcal{F}_{\k,yz}&0
   \end{array}\right).
\end{equation}

What we obtained above is the spectral BC and in order to obtain the total BC, as Bascone et. al\cite{uhlmann2019} did  we need to sum over $\k $ or integrate over $\k $ in the 1-st Brillouin zone in the limit of thermodynamics
\begin{equation}
   \mathcal{F}_{xy}=\int_{-\pi}^\pi\int_{-\pi}^\pi\mathrm{d}k_x\mathrm{d}k_y~\mathcal{F}_{\k ,xy}.\label{total}
\end{equation}

For the sake of simplicity and without loss of generality, we constrain $\{J_i\}$ with
\begin{equation}
   J_x+J_y+J_z=1\Rightarrow J_z~(J_x,J_y)=1-J_x-J_y.
\end{equation}

\begin{figure}[!htb]
   \setlength{\abovecaptionskip}{-0.6cm}
\setlength{\belowcaptionskip}{0.5cm}
   \includegraphics[scale=0.25]{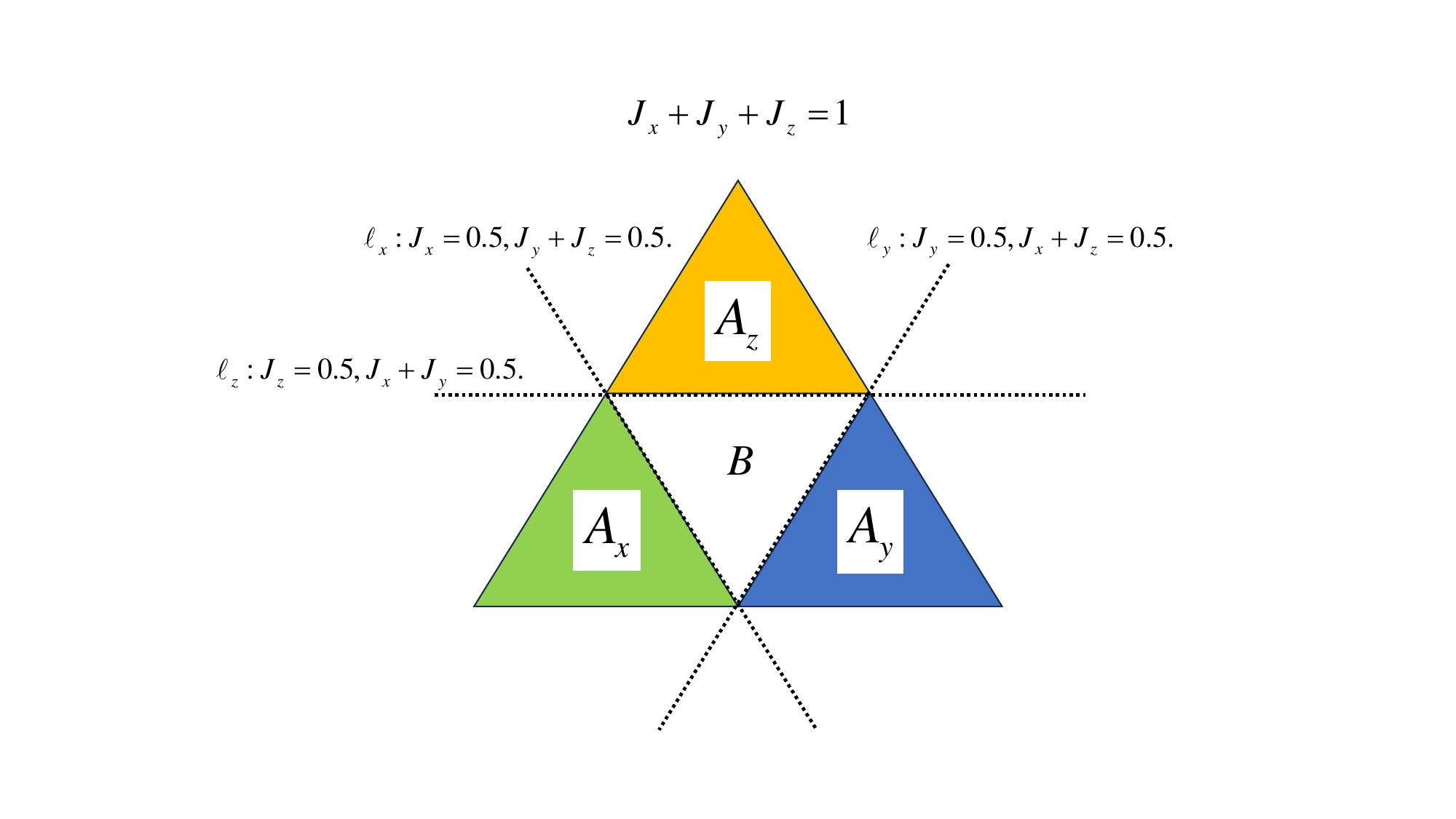}
   \centering
   \caption{The phase transition lines of Kitaev model\cite{kitaev2006}. The line $\ell_i$ represents the phase transition line from phase $A_i$ to phase $B$ under the constraint that $J_x+J_y+J_z=1$.}
   \label{transition}
\end{figure}
After this substitution, we can denote all the factors in spectral BC in terms of $J_x,J_y$,
\begin{equation}
\begin{split}
\alpha_\k &=4K(\sin k_x-\sin k_y)\\
\xi_\k &=2J_x\cos k_x+2J_y\cos k_y+2(1-J_x-J_y)\\
\beta_\k &=2J_x\sin k_x+J_y\sin k_y,
\end{split}
\end{equation}
and the spectral Berry curvatures are given by:

\begin{equation}
   \begin{split}
      \mathcal{F}_{\k ,xy}&=\frac{2J_x\cos k_x+2J_y\cos k_y+2(1-J_x-J_y)}{(\alpha_\k ^2+\beta_\k ^2+\xi_\k ^2)^{3/2}}\\
      \mathcal{F}_{\k ,yz}&=\frac{4K(\sin k_x-\sin k_y)}{(\alpha_\k ^2+\beta_\k ^2+\xi_\k ^2)^{3/2}}\\
      \mathcal{F}_{\k ,zx}&=-\frac{2J_x\sin k_x+J_y\sin k_y}{(\alpha_\k ^2+\beta_\k ^2+\xi_\k ^2)^{3/2}},
   \end{split}
\end{equation}
with
\begin{equation}
   \begin{split}
      \varepsilon_\k^3&=[16K^2(\sin k_x-\sin k_y)^2+4J_x^2\\
      &+4J_y^2+4(1-J_x-J_y)^2+8J_xJ_y\cos k_x\cos k_y\\
      &+8J_x(1-J_x-J_y)\cos k_x+8J_y\cos k_y(1-J_x-J_y)\\
      &+8J_xJ_y\sin k_x\sin k_y]^{3/2}.
   \end{split}
\end{equation}
As we mentioned before, the spectral BC was related with correlation in Eq.(\ref{susceptibility}), but for the total BC it is the summation of spectral BC or the integral on 1-st Brillouin zone in the limit of thermodynamics
so, the corresponding susceptibility needs to sum over $\k $ or integrate over $\k $ in the 1-st Brillouin zone :
\begin{equation}
   \begin{split}
      \chi_{\alpha,\beta}(\omega)&=\int_{-\pi}^\pi\int_{-\pi}^\pi\mathrm{d}^2\k \chi_{\k \alpha,\k \beta}(\omega)\\
      &=\lim_{\eta\rightarrow0}\iint \mathrm{d}\k ^2\left(\frac{\<0_\k|\sigma_{\k \alpha}|1_\k\>\<1_\k|\sigma_{\k \beta}|0_\k\>}{\omega+i\eta-(E_{\k,1}-E_{\k,0})}\right.\left.-\frac{\<0_\k|\sigma_{\k \beta}|1_\k\>\<1_\k|\sigma_{\k \alpha}|0_\k\>}{\omega+i\eta-(E_{\k,0}-E_{\k,1})}\right).
   \end{split}
\end{equation}
Then we find the relation between the total BC and the total effective spin magnetic susceptibility:
\begin{equation}
   \mathcal{F}_{\alpha,\beta}=-i\frac{\mathrm{d}}{\mathrm{d}\omega}\chi_{\alpha,\beta}(\omega)\big|_{\omega=0}, \label{totalsusceptibility}
\end{equation}
and the calculation details of the above equation is placed in Appendix A. Furthermore, we can also define the second order nonlinear magnetic susceptibility as:
\begin{equation}
   \begin{split}
      \chi_{\k,\alpha\beta\gamma}(t,t')&=(-i)^2\Theta(t-t')\Theta(t')\times\\
      &\<[[\sigma_{\k\alpha}(t),\sigma_{\k\beta}(t')],\sigma_{\k\gamma}(0)]\>_0
   \end{split}
\end{equation}
and we find the relation between the second order susceptibility and the Berry connection, for brevity we placed the result Eq.(\ref{second order}) and calculation detials in the Appendix E.

\subsection*{C. Numerical Results}
\begin{figure}[!htb]
\setlength{\belowcaptionskip}{0.5cm}
   \includegraphics[scale=0.65]{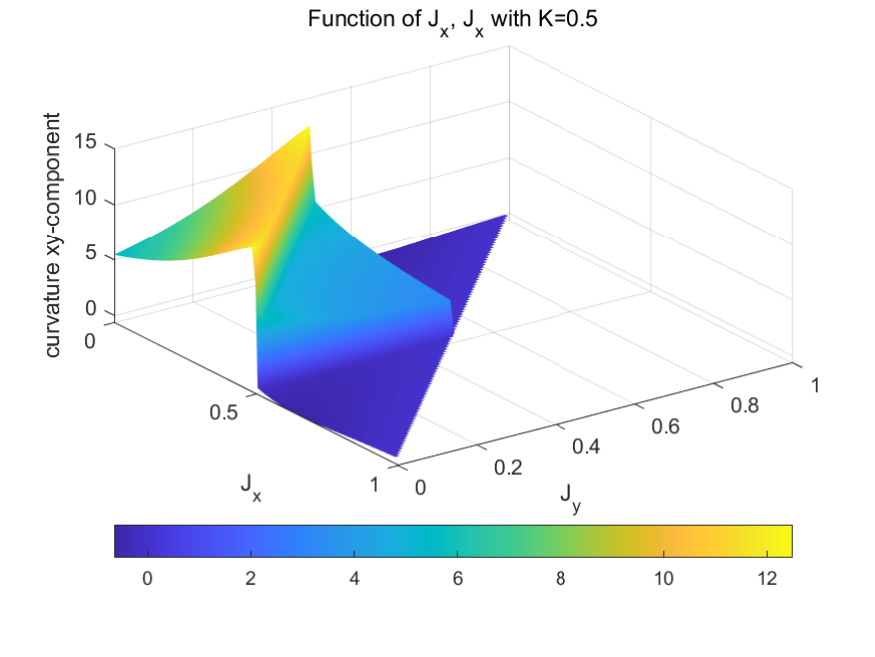}
   \setlength{\belowcaptionskip}{0.5cm}
   \caption{The xy-component of BC with constrain $J_x+J_y+J_z=1, J_x,J_y\in[0,1]$ and $K=0.5$. We can see that the surface has three stairs lines which mean the curvature's value will jump around the three lines. According to the Kitaev phase diagram Fig.\ref{transition}, they correspond the phase transition lines.}
   \centering
   \label{curvature_z}
   \label{k=0.5F_xy}

\end{figure}

\begin{figure}[!htb]
\setlength{\belowcaptionskip}{0.5cm}
   \includegraphics[scale=0.63]{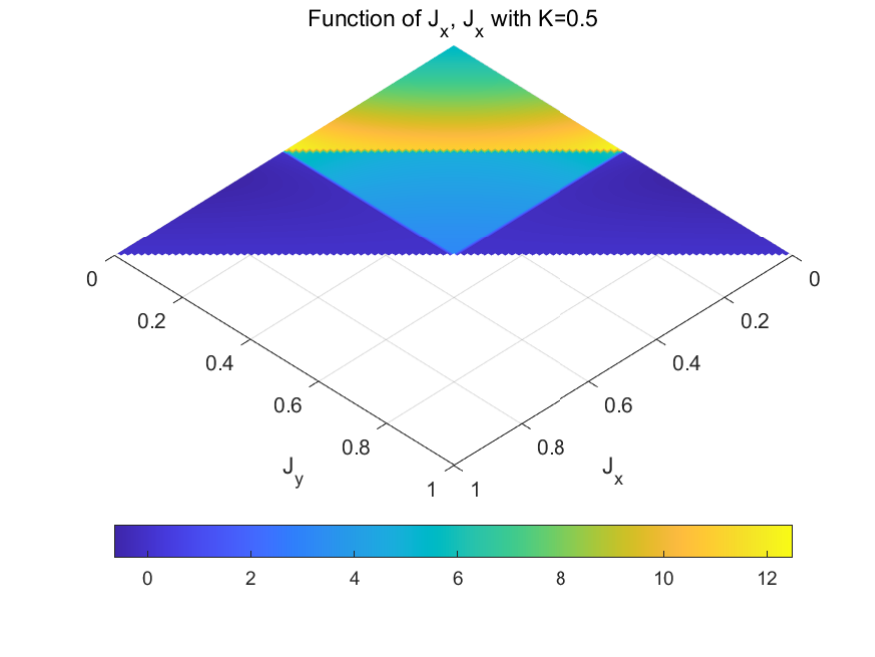}
   \setlength{\belowcaptionskip}{0.5cm}
   \centering
   \caption{The top view of Fig. \ref{k=0.5F_xy}. We can find that the top view is a regular triangle like the phase diagram Fig. \ref{transition}.}
   \label{topview}
\end{figure}

According to Eq.(\ref{total}) we do the numerical integral where $K$ the three-body interaction strength is located in $[0,1]$ and the result is shown in Fig. \ref{k=0.5F_xy} and Fig. \ref{topview}.

In Fig. \ref{topview}, the $\bigtriangledown$ (baby blue) region in the center corresponds the $B$ phase in Fig. \ref{transition} and the critical lines between the $\bigtriangledown$ region and another region correspond the phase transition lines in Fig. \ref{transition}.
If $J_x+J_y<0.5$ it means the ground state is in $A_z$ phase correspondding the topmost region of Fig.(\ref{topview}) and if $J_x+J_y>0.5$ it means the ground state will be in $B$ phase existing a phase transition from $A_z$ phase to $B$ phase which corresponds the xy-component of total BC will jump around the line $J_x+J_y=0.5,J_z=0.5$.

   If $J_y+J_z<0.5$ the ground state is in $A_x$ phase corresponding the lower-left region of Fig.(\ref{topview}) and if $J_y+J_z>0.5$ it means the ground state will be in $B$ phase existing a phase transition from $A_x$ phase to $B$ phase which corresponds the xy-component of total BC will jump around the line $J_y+J_z=0.5,J_x=0.5$.
   As we will discuss later Eq.(\ref{susceptibility}), the jumping of xy-component actually is the derivative of total effective susceptibility under the low-frequency limits.

   In other words, the signature of phase transition from $B$ phase to another $A_i$ phase is the jumping of xy-component of BC.
But as we know the BC is a geometrical quantity which is difficult to measure in experiments so, we've related this to the effective spin magnetic susceptibility Eq.(\ref{totalsusceptibility}), i.e., the derivative of total effective spin magnetic susceptibility has the jumping for the transition from $B$ phase to another phases.

\begin{figure*}[htbp]
    \centering
    \begin{subfigure}[b]{0.45\textwidth}
        \centering
        \includegraphics[width=\textwidth]{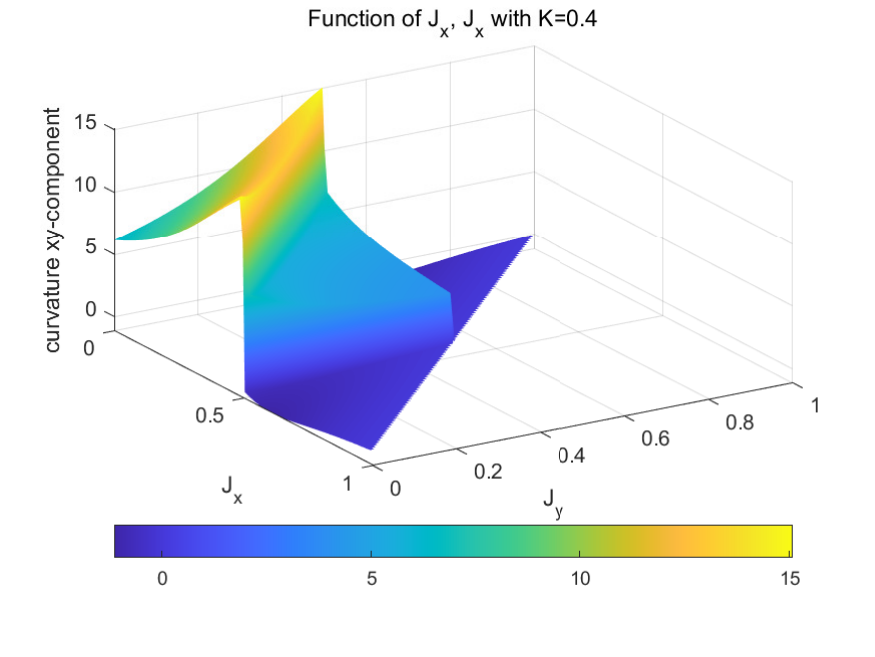}
        \caption{xy-component of BC with $K$=0.4}
        \label{k=0.4}
    \end{subfigure}
    \hfill
    \begin{subfigure}[b]{0.45\textwidth}
        \centering
        \includegraphics[width=\textwidth]{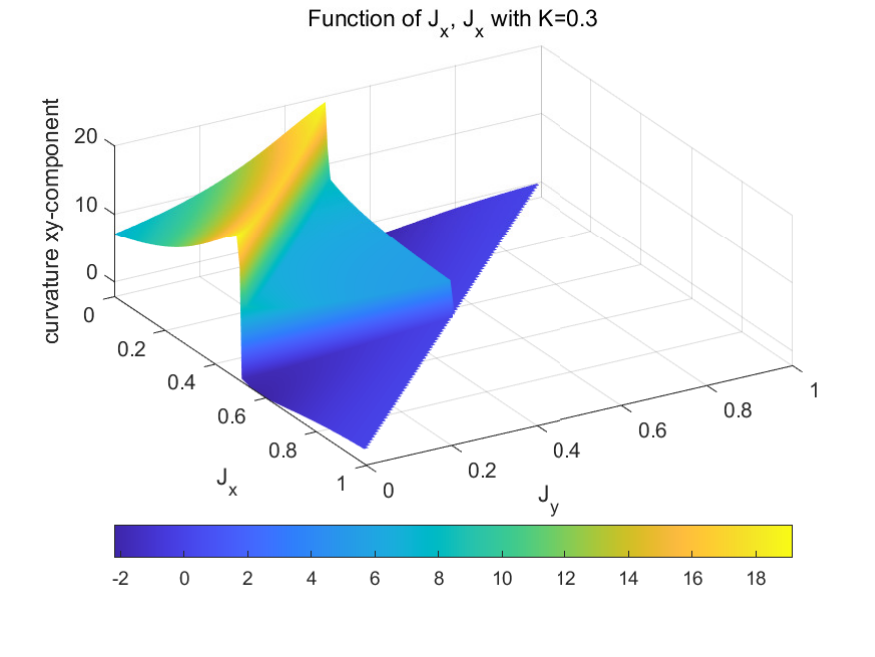}
        \caption{xy-component of BC with $K$=0.3}
        \label{k=0.3}
    \end{subfigure}
    \vskip\baselineskip 

    \begin{subfigure}[b]{0.45\textwidth}
        \centering
        \includegraphics[width=\textwidth]{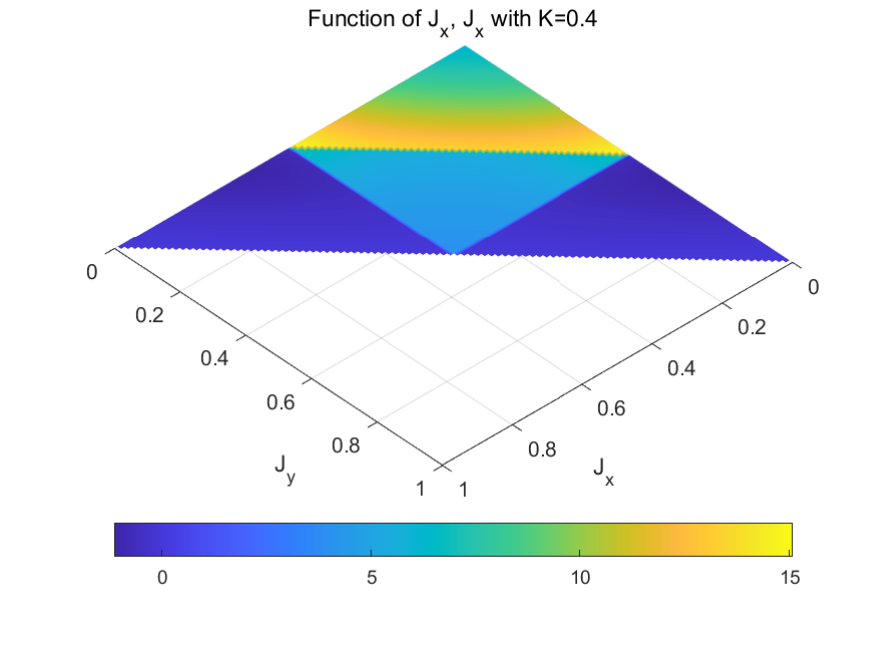}
        \caption{Top view of xy-component of BC with $K$=0.4}
        \label{topviewk=0.4}
    \end{subfigure}
    \hfill
    \begin{subfigure}[b]{0.45\textwidth}
        \centering
        \includegraphics[width=\textwidth]{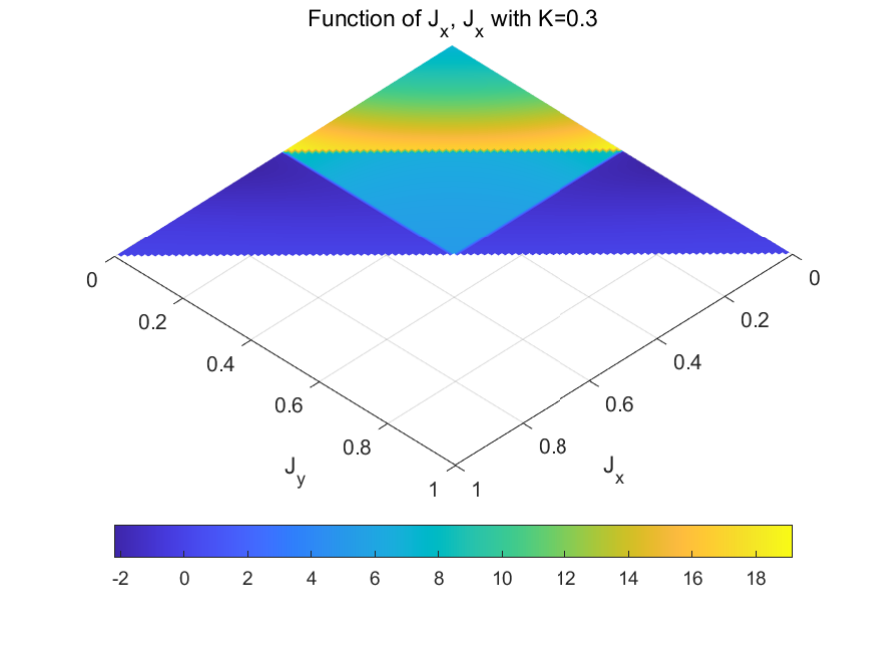}
        \caption{Top view of xy-component of BC with $K$=0.3}
        \label{topviewk=0.3}
    \end{subfigure}

    \caption{BC with different $K$ values.}
    \label{k=0.40.3}
\end{figure*}

   In addition, we find that the curvature's property characterizing the phase transition will not change with the different $K$, i.e., the critical lines will not change with the changing of $K$, as shown in Fig.~\ref{k=0.40.3}.
   In other words, we claim the topological phase transition has the robustness against the perturbation.
   In concluding, we related the xy-component BC with the total effective susceptibility and it reveals the topological critical behavior around the phase transition lines.

\section*{III. Berry curvature with finite temperature-Uhlmann curvature}

\begin{figure*}[htbp]
   \centering
   \begin{subfigure}[b]{0.45\textwidth}
       \centering
       \includegraphics[scale=0.63]{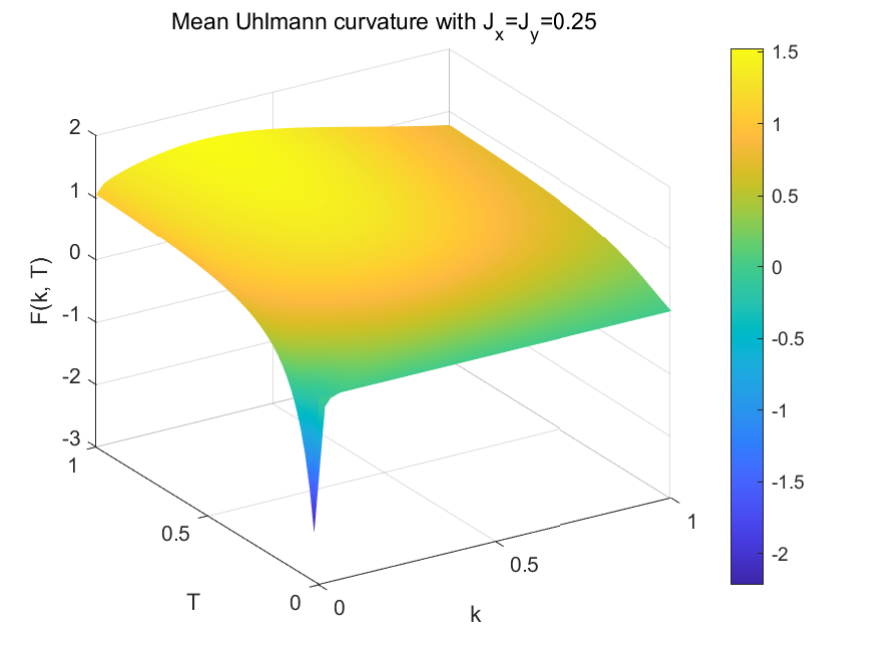}
       \caption{In this Figure, we set $J_x=J_y=0.25,J_z=0.5$ at the phase transition point. And the Uhlmann curvature peaked at $T=0,K=0$.}
       \label{MU}
   \end{subfigure}
   \hfill
   \begin{subfigure}[b]{0.45\textwidth}
       \centering
       \includegraphics[scale=0.63]{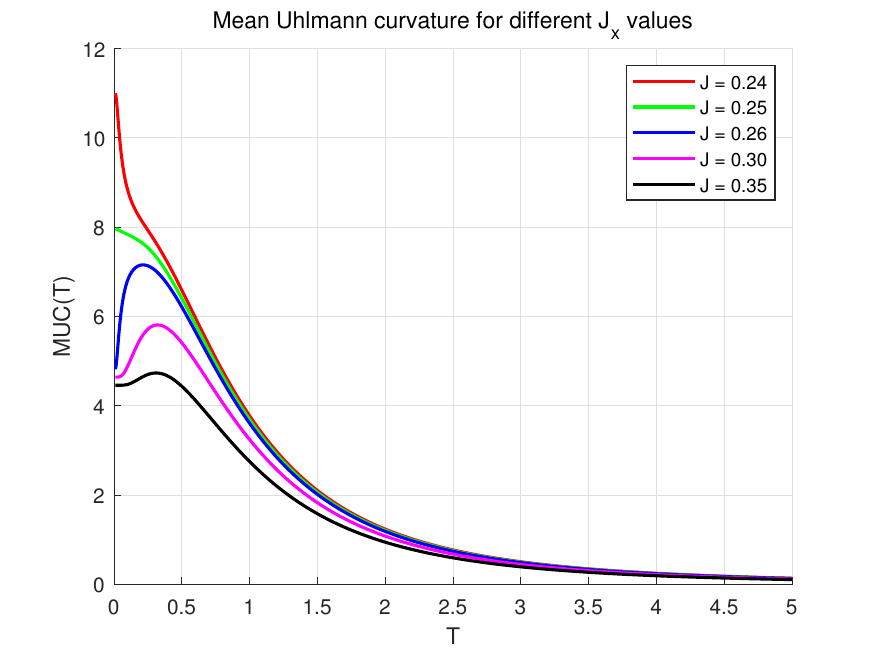}
       \caption{This figure shows that the xy-component of the mean Uhlmann curvature changing with temperature $T$ assuming $J_y=0.25, J_z=1-J_x-J_y,K=0.5$.}
       \label{differentJ_x}
   \end{subfigure}
   \vskip\baselineskip 
   \caption{xy-component of mean Ulhmann curvature}
\end{figure*}

   What we analysed above is based on the zero-temperature situation, now we try to generalize it to the finite-temperature system\cite{dissipative,invariantsystem}. Uhlmann was the first to propose the geometric phase for the mixed state and generalize the BC to finite-temperature situation. 
   The core idea of Uhlmann paradigm is purification. According to this paradigm, any mixed state density operator can be regarded as the reduced density operator of a pure state lying in an enlarged Hilbert space\cite{QPT}.
     In 2019, Wang proposed a sub-geometric phase which is defined by density matrix and he generalized the geometric phase to the mixed state too\cite{subgeometric}. According to\cite{dissipative,invariantsystem} the Uhlmann curvature can be simplified as
\begin{equation}
   \mathcal{U}_{\k,\alpha\beta}=\frac{i}{\pi}\int_{-\infty}^\infty\frac{\mathrm{d}\omega}{\omega^2}\tanh^2\left(\frac{\omega}{2T}\right)\int_{-\infty}^\infty\mathrm{d}t\mathbf{e}^{i\omega t}\zeta_{\k,\alpha\beta}(t),
\end{equation}
with
\begin{equation}
   \zeta_{\alpha\beta}(t)=\frac12\<[\d_\alpha H(t),\d_\beta H(0)]\>_0.
\end{equation}
Similarly, we sum over $\k$ in the 1-st Brillouin zone and obtain the total mean Uhlmann curvature:
\begin{equation}
   \mathcal{U}_{\alpha\beta}=\int\mathrm{d}\k^2 ~\mathcal{U}_{\k,\alpha\beta}.
\end{equation}
In the low temperature limits as $T\rightarrow0$, the Uhlmann curvature will degenerate to the BC.
Then we set $J_x=J_y=0.25, J_z=0.5$ and the Figure that Uhlmann curvature changing with $K$ and $T$ is shown in Fig. \ref{MU}. In Fig.(\ref{MU}) we can see that there is a peak at $K=0$. It's because if we don't break the time-reversal symmetry and the curvature will be vanished, in other words it reveals that the local perturbation will open a gap existing the non-Abelian anyon excitation.

   In Fig.(\ref{differentJ_x}) we find that when $J_x=0.24$ the ground state will be in $A_z$ phase because of the constraint $J_y=0.25$ and the state will be on the critical phase transition line from $A_z$ phase to $B$ phase if $J_x=0.25$ . If $J_x$ is increased to $J_x=0.26$ or $J_x>0.25$ the curve will have an extrema at about $T=0.3$.
   Similarly, if we adjust $J_y$ and $J_z$ it will reveal the phase transition from $A_x$ phase to $B$ phase instead.

As we mentioned before, the curvature has robustness against local perturbation. We can see that if we adjust $K$ in different values, the peak will not move largely at $T=0.3$.
According to \cite{invariantsystem}, the Uhlmann curvature can be written by the form of:
\begin{equation}
   \mathcal{U}_{\k,\alpha\beta}=\frac{i}{\pi}\int\frac{\mathrm{d}\omega}{\omega^2}\tanh^2\left(\frac{\omega}{2T}\right)\zeta_{\k,\alpha\beta}(\omega),\label{Uhlmann and spectral}
\end{equation}
where $\zeta(\omega)$ is the spectral function of the correlation and it is given by:
\begin{equation}
   \begin{split}
         \zeta_{\k,\alpha\beta}(\omega)&=\frac12\int\mathrm{d}t \mathbf{e}^{i\omega t}\<[\sigma_{\k,\alpha}(t),\sigma_{\k,\beta}(0)]\>,\label{spectral}
   \end{split}
\end{equation}
and we substitute the implicit expression of $\zeta_{\k,\alpha\beta}$ into the mean Uhlmann curvature:

\begin{equation}
   \begin{split}
      \mathcal{U}_{\k,\alpha\beta}&=\tanh^2\left(\frac{\varepsilon_\k}{T}\right)\tanh\left(\frac{\varepsilon_\k}{2T}\right)\mathcal{F}_{\k,\alpha\beta}.\label{Berry and Uhlmann}
   \end{split}
\end{equation}
We can see if $T\to0^+$ then the above equation will converge to the Berry curvature, i.e. the pure state Berry curvature;
\begin{equation}
   \lim_{T\to0^+}\mathcal{U}_{\k,\alpha\beta}=\mathcal{F}_{\k,\alpha\beta}.
\end{equation}
We can see the mixed state Uhlmann curvature Eq.(\ref{Uhlmann and spectral}) and the pure state Berry curvature Eq.(\ref{susceptibility}) are different in form but we can substitute Eq.\eqref{Berry and Uhlmann} into Eq.\eqref{susceptibility} then we'll find that

\begin{equation}
   \mathcal{U}_{\k,\alpha\beta}=-i\tanh^2\left(\frac{\varepsilon_\k}{T}\right)\tanh\left(\frac{\varepsilon_\k}{2T}\right)\frac{\mathrm{d}}{\mathrm{d}\omega}\chi_{\k,\alpha\beta}\big|_{\omega=0}.
\end{equation}

\section*{IV. Quantum geometric tensor and Fubini-Study metric}
\subsection*{A. Zero temperature
}
In 1980s~\cite{chengran}, many physists discovered that the gauge field can appear in some simple quantum system with some conditions which resulted the emergence of metric structure in the quantum system.
The Fubini-Study metric actually is the metric tensor of the Hilbert space deriving from the quantum distance between different states\cite{quantumdistance}. We consider a Hamiltonian $\{H(\lambda_i)\}$ changing with a set of parameters $\lambda_i$, the quantum geometric tensor is given by\cite{chengran}:

\begin{equation}
   \begin{split}
      Q_{\alpha\beta}(\lambda)&=\<\d_\alpha\psi(\lambda)|1-\mathcal{\hat P}|\d_\beta\psi(\lambda)\>\\
      &=\<\d_\alpha\psi|\d_\beta\psi\>-\<\d_\alpha\psi|\psi\>\<\psi|\d_\beta\psi\>,
   \end{split}
\end{equation}
where $\mathcal{\hat P}$ is the projective operator $\mathcal{P}=|\psi\>\<\psi|$. The BC is directly related with the imaginary part of $Q$ with a coefficient
\begin{equation}
   \mathcal{F}_{\alpha\beta}=-2\mathbf{Im}~ [Q_{\alpha\beta}]=i(Q_{\alpha\beta}-Q_{\beta\alpha}),
\end{equation}
it has the same form as Eq.(\ref{curvature}). The Fubini-Study metric is defined as:
\begin{equation}
   g_{\alpha\beta}:=\mathbf{Re}~[Q_{\alpha\beta}]=\frac12(Q_{\alpha\beta}+Q_{\beta\alpha}),\label{definemetric}
\end{equation}
and the geometric tensor can be rewritten as:
\begin{equation}
   Q_{\alpha\beta}=g_{\alpha\beta}-\frac{i}{2}\mathcal{F}_{\alpha\beta},
\end{equation}
similarly we insert a set of complete orthonormal basis in $Q$ and use the identity Eq.(\ref{identity}) then we can simplify it into
\begin{equation}
   Q_{\alpha\beta}=\sum_{n\neq0}\frac{\<\psi_0|\d_\alpha H|\psi_n\>\<\psi_n|\d_\beta H|\psi_0\>}{(E_0-E_n)^2}.\label{definition of geometric tensor}
\end{equation}
It's easy to calculate the $\k$-fixed components of the geometric tensor with $\k$-dependent Hamiltonian Eq.(\ref{effectivemagnetic}) by:
\begin{equation}
   \begin{split}
      Q_{\k,\alpha\beta}&=\frac{\<0_\k|\sigma_{\k\alpha}|1_\k\>\<1_\k|\sigma_{\k\beta}|0_\k\>}{4\varepsilon_\k^2}.\label{tensor}
   \end{split}
\end{equation}
In order to facilitate the calculation, we'll parameterize the effective magnetic field Eq.(\ref{B_k}) into the spherical coordinates by:
\begin{equation}
   \mathbf{B}_\k=|\mathbf{B}_\k|(\sin\theta_\k\cos\phi_\k,\sin\theta_\k\sin\phi_\k,\cos\theta_\k),
\end{equation}
where these parameters are defined as:
\begin{equation}
   \begin{split}
      \theta_\k&=\arccos\left(\frac{B_z}{\varepsilon_\k}\right)=\arccos\left(\frac{\xi_\k}{\sqrt{\alpha_\k^2+\beta_\k^2+\xi_\k^2}}\right)  \\\\
      \phi_\k&=\arctan\left(\frac{B_y}{B_x}\right)=\arctan\left(\frac{-\xi_\k}{\alpha_\k}\right)\\\\
      |\mathbf{B}_k|&=\varepsilon_\k=\sqrt{\alpha_\k^2+\beta_\k^2+\xi_\k^2}.\label{parameterization}
   \end{split}
\end{equation}
Under this parameterization, the Hamiltonian Eq.(\ref{effectivemagnetic}) can be rewritten as:
\begin{equation}
   H(\k)=\varepsilon_\k\left(
      \begin{array}{cc}
         \cos\theta_\k&\sin\theta_\k \mathbf{e}^{-i\phi_\k}\\\\
         \sin\theta_\k \mathbf{e}^{i\phi_\k}&-\cos\theta_\k
      \end{array}
   \right),
\end{equation}
where we can find the two eigenstates in many quantum textbooks as:
\begin{equation}
      |0_\k\>=\left(\begin{array}{cc}
         -\sin\frac{\theta_\k}{2}\mathbf{e}^{-i\frac{\phi_\k}{2}}\\\\
         \cos\frac{\theta_\k}{2}\mathbf{e}^{i\frac{\phi_\k}{2}}
      \end{array}\right)\label{ground state}
   \end{equation}
   
\begin{equation}
      |1_\k\>=\left(
         \begin{array}{cc}
            \cos\frac{\theta_\k}{2}\mathbf{e}^{-i\frac{\phi_\k}{2}}\\\\
            \sin\frac{\theta_\k}{2}\mathbf{e}^{i\frac{\phi_\k}{2}}
         \end{array}
      \right),\label{excited state}
\end{equation}
then we substitute the above two eigenstates into Eq.(\ref{tensor}) and the results are placed in Appendix D.

\begin{figure*}[htbp]
   \centering
   \begin{subfigure}[b]{0.45\textwidth}
       \centering
       \includegraphics[scale=0.63]{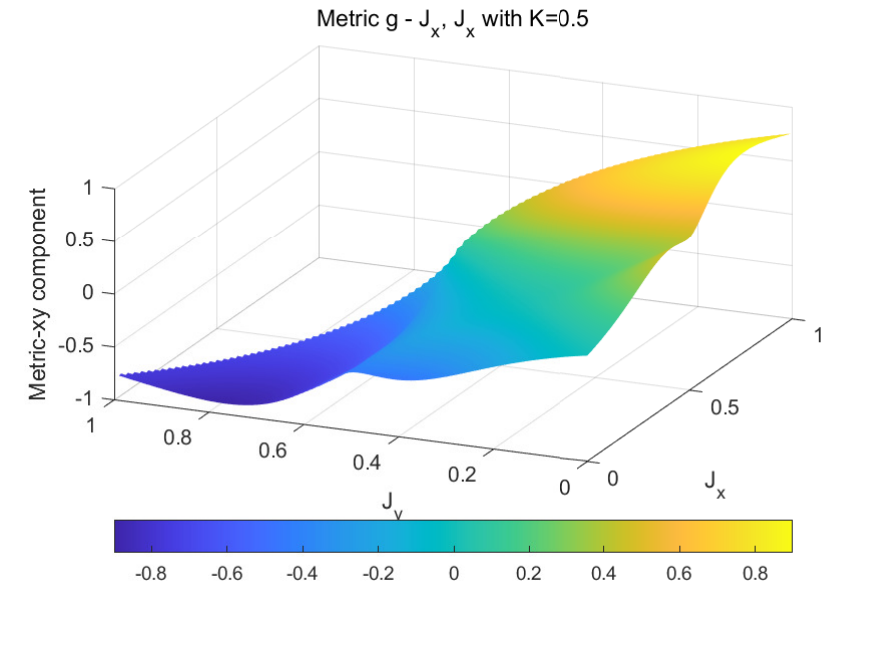}
       \caption{The xy-component of Fubini-Study metric with fixed $K=0.5$.}
       \label{metric_xy}
   \end{subfigure}
   \hfill
   \begin{subfigure}[b]{0.45\textwidth}
       \centering
       \includegraphics[scale=0.63]{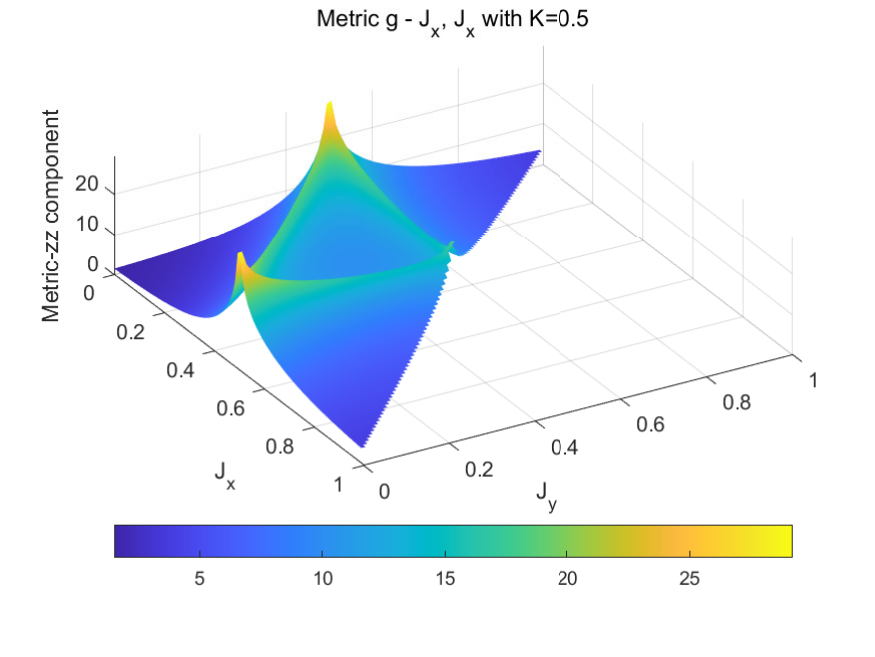}
       \caption{The zz-component of Fubini-Study metric with fixed $K=0.5$.}
       \label{metric_zz}
   \end{subfigure}
   \vskip\baselineskip 
   \caption{Different components of Fubini-Study metric with $K=0.5$.}
\label{metric_xy,zz}
\end{figure*}

Hence, the quantum geometric tensor can be represented by a matrix:
\begin{equation}
   \mathbf{Q}_\k=\left(\begin{array}{ccc}
      Q_{\k,xx}&Q_{\k,xy}&Q_{\k,xz}\\\\
      Q_{\k,xy}^*&Q_{\k,yy}&Q_{\k,yz}\\\\
      Q_{\k,xz}^*&Q_{\k,yz}^*&Q_{\k,zz}
   \end{array}\right),
\end{equation}
and the Fubini-Study metric Eq.(\ref{definemetric}) is the real symmetric matrix. According to Kolodrubetz et al.'s work\cite{geometry2017}, the metric tensor can be written as:
\begin{equation}
   g_{\k,\alpha\beta}=\int_0^\infty\frac{\mathrm{d}\omega}{2\pi}\frac{\mathbf{Im}\chi_{\k,\alpha\beta}(\omega)+\mathbf{Im}\chi_{\k,\beta\alpha}(\omega)}{\omega^2}.\label{55}
\end{equation}
Because what we obtained is $\k$-fixed in order to obtain the total Fubini-Study metric we need to
sum over $\k$ or integral over $\k$ in the 1-st Brillouin zone under the thermodynamics limit:
\begin{equation}
   g_{\alpha\beta}=\iint \mathrm{d}\k^2 g_{\k,\alpha\beta}.
\end{equation}
Then we can calculate the xy-component and zz-component of Fubini-Study metric numerically with fixed interaction strength $K=0.5$ and the Figures are shown as Fig.(\ref{metric_xy},\ref{metric_zz}).
As we can see, Fig.(\ref{metric_zz}) has the similar behavior as the xy-component of BC in Fig.(\ref{curvature_z}) which has the jumping around the phase transition lines.

\subsection*{B. Bures metric of mixed state}

Like the mean Uhlmann curvature, one can generalize the Fubini-Study metric of pure state to the Bures metric of mixed state, i.e. the metric with finite temperature\cite{Bures,Bures1969}. Consider a general Hamiltonian $H(\lambda_i)$ depending on $\{\lambda_i\}$ and we use $\partial_\alpha$ to denote the derivative with respect to $\lambda_\alpha$. 
The Bures metric is defined as\cite{QPT,Bures,Bures1969}
\begin{equation}
   \mathcal{G}_{\alpha\beta}=\frac14\Tr\left[\rho\{L_\alpha,L_\beta\}\right],
\end{equation}
where $\{,\}$ represents the fermionic commutator and $L_\mu$ is the symmetric logarithmic derivative(SLD) defined by this following equation\cite{QPT}
\begin{equation}
   \partial_{\alpha}\rho=\frac12\{\rho,L_\alpha\},\label{SLD}
\end{equation}
here $\rho$ is the denstiy matrix operator. And the geometric tensor for mixed state is defined as\cite{QPT}
\begin{equation}
   \mathcal{Q}_{\alpha\beta}=\frac14\Tr[\rho L_\alpha L_\beta].
\end{equation}
Based on the form of the quantum geometric tensor for mixed states, it is evident that the Bures metric is actually the symmetric part of the geometric tensor, while the Uhlmann curvature corresponds to its antisymmetric part:
\begin{equation}
   \mathcal{U}_{\alpha\beta}=\frac i4\mathbf\Tr[\rho[L_\alpha,L_\beta]],
\end{equation}
as we have calculated in Eq.(\ref{Uhlmann and spectral}). The explicit form of the Bures metric can be easily computed \cite{Bures1969,QPT}
\begin{equation}
   \begin{split}
      \mathcal{G}_{\alpha\beta}&=\frac12\sum_i\frac{\d_\alpha p_i\d_\beta p_i}{p_i}-\sum_{i,j}\frac{(p_i-p_j)^2}{p_i+p_j}\<i|\d_\alpha|j\>\<j|\d_\beta|i\>\\
      &=\frac12\sum_i\frac{\d_\alpha p_i\d_\beta p_i}{p_i}-\sum_{i,j}\frac{(p_i-p_j)^2}{p_i+p_j}\frac{\<i|\d_\alpha H|j\>\<j|\d_\beta H|i\>}{(E_i-E_j)^2},
      \label{Bures metric}
   \end{split}
\end{equation}
where $p_i$ is the probability in Gibbs ensemble defined as:
\begin{equation}
   p_i=\frac{\mathbf{e}^{-E_i/T}}{\sum_{j}\mathbf{e}^{-E_j/T}}.
\end{equation}
It can be observed that the first term $\mathcal{G}^{(1)}_{\alpha\beta}$ in Eq.(\ref{Bures metric}) is an additional component compared to the metric for pure state, which actually corresponds to the Rao-Fisher metric\cite{QPT,Bures1969} instead, the second term $\mathcal{G}^{(2)}_{\alpha\beta}$ corresponds the Fubini-Study metric for pure state.
We can use the following mathematical identity\cite{geometry2017}
\begin{equation}
   \frac{p_i-p_j}{p_i+p_j}=\int\mathrm{d}\omega~\tanh\left(\frac{\omega}{2T}\right)\delta(\omega-E_j+E_i),
\end{equation}
and substitute it into the second term $\mathcal{G}^{(2)}_{\alpha\beta}$ of the Bures metric Eq.(\ref{Bures metric}) and we can find that
\begin{equation}
   \begin{split}
      \mathcal{G}^{(2)}_{\alpha\beta}&=\frac{1}{\pi}\int_{-\infty}^\infty\frac{\mathrm{d}\omega}{\omega^2}~\tanh\left(\frac{\omega}{2T}\right)\zeta_{\alpha\beta}(\omega).\label{G2}
   \end{split}
\end{equation}
We use $G(\omega,T)$ to denote the integrand of the above and $F(\omega,T)$ to denote the integrand of Eq.(\ref{Uhlmann and spectral}), respectively and we find they satisfy an easy relation 
\begin{equation}
   F(\omega,T)=i\tanh\left(\frac{\omega}{2T}\right)G(\omega,T).
\end{equation}
Then we use the above relation into the model we studied which is a two-band model with two $\k$-dependent eigen-energies $\pm\varepsilon_\k$ and we'll obtain
\begin{equation}
   \begin{split}
      \mathcal{G}^{(2)}_{\k,\alpha\beta}&=\frac{1}{\pi}\int_{-\infty}^\infty\frac{\mathrm{d}\omega}{\omega^2}~\tanh\left(\frac{\omega}{2T}\right)\zeta_{\k,\alpha\beta}(\omega)\\
      &=\frac12\tanh^2\left(\frac{\varepsilon_\k}{T}\right)(Q_{\k,\alpha\beta}+Q_{\k,\alpha\beta}^*)\\
      &=\tanh^2\left(\frac{\varepsilon_\k}{T}\right)g_{\k,\alpha\beta}.
   \end{split}
\end{equation}
Here, $g_{\k,\alpha\beta}$ is defined as Eq.(\ref{definemetric}) and we can see if temperature $T\to 0^+$ the second term of Bures metric $\mathcal{G}^{(2)}_{\alpha\beta}$ will converge to the Fubini-Study metric.
If we differentiate $\mathcal{G}^{(2)}$ with respect to $T$ then we'll find
\begin{equation}
   \begin{split}
      \frac{\mathrm{d}\mathcal{G}^{(2)}_{\k,\alpha\beta}}{\mathrm{d}T}=&-\frac{1}{2\pi T^2}\int\frac{\mathrm{d}\omega}{\omega}\zeta_{\k,\alpha\beta}(\omega)\\
      &+\frac{1}{2\pi T^2}\int\frac{\mathrm{d}\omega}{\omega}\tanh^2\left(\frac{\omega}{2T}\right)\zeta_{\k,\alpha\beta}(\omega).
   \end{split}
\end{equation}
according to Eq.\eqref{spectral representation}, we can simplicity the above equation into:
\begin{equation}
   \frac{\mathrm{d}\mathcal{G}^{(2)}_{\k,\alpha\beta}}{\mathrm{d}T}=-\frac{1}{2T^2}\left(\chi_{\k,\alpha\beta}(0)-\int\frac{\mathrm{d}\omega}{\omega}\tanh^2\left(\frac{\omega}{2T}\right)\zeta_{\k,\alpha\beta}(\omega)\right)
\end{equation}
Besides, we can see that the Fisher-Rao metric $\mathcal{G}^{(1)}_{\alpha\beta}$ actually arises due to the mixed state and we'll demonstrate that as $T\to 0^+$ this term will vanish\cite{QPT}. 

The Fisher-Rao metric $\mathcal{G}^{(1)}_{\alpha\beta}$ can be calculated as:
\begin{equation}
   \begin{split}
      \mathcal{G}^{(1)}_{\k,\alpha\beta}&=\frac12\left(\frac{\d_{\k,\alpha}p_0\d_{\k,\beta}p_0}{p_0}+\frac{\d_{\k,\alpha}p_1\d_{\k,\beta}p_1}{p_1}\right)\\
      &=\frac12\frac{B_{\k,\alpha}B_{\k,\beta}}{\varepsilon_\k^2T^2\cosh^2\left(\frac{\varepsilon_\k}{T}\right)} \stackrel{T \rightarrow 0^+}{=} 0.
   \end{split}
\end{equation}
As we predicted before, as temperature $T\to 0^+$ the Fisher-Rao metric term will vanish. In that case, we can write the Bures metric as:
\begin{equation}
   \begin{split}
      \mathcal{G}_{\k,\alpha\beta}&=\tanh^2\left(\frac{\varepsilon_\k}{T}\right)g_{\k,\alpha\beta}+\frac12\frac{B_{\k,\alpha}B_{\k,\beta}}{\varepsilon_\k^2T^2\cosh^2\left(\frac{\varepsilon_\k}{T}\right)}.
   \end{split}
\end{equation}
 Hence, we can easily write the geometric tensor at finite temperature for mixed state:
\begin{equation}
   \begin{split}
      \mathcal{Q}_{\k,\alpha\beta}&=\frac14\Tr[\rho\{L_\alpha,L_\beta\}]=\mathcal{G}_{\k,\alpha\beta}-\frac i2\mathcal{U}_{\k,\alpha\beta}.
   \end{split}
\end{equation} 

\section*{V. Conclusion}
In this paper, we derived the BC in effective magnetic field parameter space and simplified the Hamiltonian into a closed form in order to map it into the spin in effective magnetic field. In this parameter space we calculated the spectral BC with fixed $\k$ and did the integral over the 1-st Brillouin zone under the thermodynamics limit to obtain the total BC. 
   According to \cite{spinberry}, we related the total BC and the effective susceptibility as Eq.(\ref{susceptibility}). Then we numerically calculated the xy-component of total BC with constrain $J_x+J_y+J_z=1$ and found that the xy-component of total BC is highly similar to the phase diagram of Kitaev spin liquid. In each transition critical line from $A_i$ phase to $B$ phase, the xy-component of total BC has a jumping shown in Fig.(\ref{curvature_z}). At the end of this section, we derived the second nonlinear susceptibility which can be related with the non-Abelian Berry connection.

      Besides, we also found that the behavior of xy-component of total BC will not be influenced by the local perturbation, i.e. the behavior of xy-component of total BC around the critical line will be the same with different $K$ values as shown in Fig.(\ref{k=0.40.3}) which can be regarded as the signature about the topological phase transition of Kitaev spin liquid. In short, the ground state of Kitaev spin liquid has the robustness against the local perturbation which is also the property of topological order.
      Theoretically, the phase transition of Kitaev spin liquid can be described by the effective susceptibility as defined Eq.(\ref{susceptibility}). In other words, the derivative of susceptibility at $\omega=0$ will jump around the critical transition lines.

      In Sec.~IV we generalized the BC to the mixed state situation in the finite temperature inspired by \cite{uhlmann2019,invariantsystem,dissipative}. Here we still chose the effective mangetic field as the parameter, first we draw the figure Fig. (\ref{MU}) of mean Uhlmann curvature changing with $T$ and $K$ and Fig.(\ref{differentJ_x}) that Uhlmann curvature changes with $T$ with diifferent $J_x$ in which one can see there will be the phase transition from $A_z$ phase to $B$ phase with arising an extrema.
      Then we analytically calculated the mean Uhlmann curvature related to the spectral function and find the relation with the Berry curvature besides, we related the spin density correlation with the mean Uhlmann curvature by means of the spectral representation of the Green function.

   The quantum geometric tensor has the generous physical meaning whose imaginary part is the BC we calculated before and real part is the Fubini-Study metric. In Sec.IV, we analytically calculated the geometric tensor and all components of Fubini-Study metric by parameterizing the effective magnetic field into the spherical coordinates as Eq.(\ref{explicit form of tensor}).
   This metric characterizes the geometry of the parameter space where we calculated the zz-component and the xy-component of it as Shown in Fig.(\ref{metric_xy,zz}). We can see that the surface of Fig.(\ref{metric_zz}) has the same behavior as xy-component of the total BC Fig.(\ref{curvature_z}) and will peak at every cross points of any two critical lines. 
   Like Sec. IV, in B part, we generalized the Fubini-Study metric to the Bures metric for the mixed state which has an additional term called Fisher-Rao metric caused by the mixed state. And the second term will converge to the Fubini-Study metric as $T\to0^+$.
   
   \section*{Acknowledgements}
   This study is supported by the National Key R\&D Program of China (Grant No. 2022YFA1402703). We also thank Prof. Gang Su and Prof. Bo Gu for their helpful disscussions.
\newpage

\newpage

\begin{widetext}
   \section*{Appendix}
\subsection*{A. Details of linear susceptibility calculation}
In this part, we'll show the details of deriving the linear effective susceptibility Eq.(\ref{FT of susceptibility}) and the second-order nonlinear susceptibility. According to the definition of susceptibility Eq.(\ref{definition of susceptibility}), we open the commutator under the Heisenberg picture:
\begin{equation}
   \begin{split}
      \chi_{\k,\alpha\beta}(t)&=-i\Theta(t)\<[\sigma_{\k \alpha}(t),\sigma_{\k\beta}(0)]\>_0\\
      &=-i\Theta(t)\<0_\k|\sigma_{\k \alpha}(t)\sigma_{\k\beta}(0)-\sigma_{\k\beta}(0)\sigma_{\k \alpha}(t)|0_\k\>\label{details of susceptibility}
   \end{split}
\end{equation}

with
\begin{equation}
\sigma_{\k \alpha}(t)=\mathbf{e}^{iH_\k t}\sigma_{\k \alpha}(0)\mathbf{e}^{-iH_\k t}.\label{Heisenberg}
\end{equation}
Then we substitute the above relation into Eq.(\ref{details of susceptibility}) and insert a set of complete orthonormal basis and for the sake of simplicity, we denote $\sigma_{\k\alpha}(0)$ by $\sigma_{\k\alpha}$:
\begin{equation}
   \begin{split}
      \chi_{\k,\alpha\beta}(t)&=-i\Theta(t)\<0_\k|\mathbf{e}^{iH_\k t}\sigma_{\k \alpha}\mathbf{e}^{-iH_\k t}|1_\k\>\<1_\k|\sigma_{\k\beta}|0_\k\>-\<0_\k|\sigma_{\k\beta}|1_\k\>\<1_\k|\mathbf{e}^{iH_\k t}\sigma_{\k\alpha}\mathbf{e}^{-iH_\k t}|0_\k\>\\
      &=-i\Theta(t) \mathbf{e}^{i(E_{\k,0}-E_{\k,1})t}\<0_\k|\sigma_{\k\alpha}|1_\k\>\<1_\k|\sigma_{\k\beta}|0_\k\>-\mathbf{e}^{-i(E_{\k,0}-E_{\k,1})t}\<0_\k|\sigma_{\k\beta}|1_\k\>\<1_\k|\sigma_{\k\alpha}|0_\k\>
   \end{split}
\end{equation}
Then we Fourier transform it from time domain into frequency domain:
\begin{equation}
   \begin{split}
      \chi_{\k,\alpha\beta}(\omega)=\lim_{\eta\rightarrow 0^+}\frac{1}{2\pi}\iint\mathrm{d}x\mathrm{d}t&\times\left(\frac{\mathbf{e}^{i(E_{\k,0}-E_{\k,1})t}\mathbf{e}^{-ixt}}{x+i\eta}\<0_\k|\sigma_{\k\alpha}|1_\k\>\<1_\k|\sigma_{\k\beta}|0_\k\>\mathbf{e}^{i\omega t}\right.\\
      &\left.-\frac{\mathbf{e}^{-i(E_{\k,0}-E_{\k,1})t}\mathbf{e}^{-ixt}}{x+i\eta}\<0_\k|\sigma_{\k\beta}|1_\k\>\<1_\k|\sigma_{\k\alpha}|0_\k\>\mathbf{e}^{i\omega t}\right),
   \end{split}
\end{equation}
where we used the integral expression of step function $\Theta(t)$
\begin{equation}
   \Theta(t)=-\frac{1}{2\pi i}\lim_{\eta\rightarrow 0^+}\int_{-\infty}^{+\infty}\mathrm{d}x\frac{\mathbf{e}^{-ixt}}{x+i\eta}.
\end{equation}
By means of the integral expression of Dirac $\delta$ function
\begin{equation}
   \frac{1}{2\pi}\int_{-\infty}^{+\infty}\mathrm{d}t \mathbf{e}^{it(E_{\k,0}-E_{\k,1}-x+\omega)}=\delta(E_{\k,0}-E_{\k,1}-x+\omega),
\end{equation}

   finally we obtain the correlation function in terms of Eq.(\ref{FT of susceptibility}).

\subsection*{B. The relation between total BC and spectral BC}
The spectral Berry curvature \cite{uhlmann2019} is defined as:
\begin{equation}
   \mathcal{F}_{\k,\alpha\beta}=-2\textbf{Im}[\<\d_{\k\alpha}0_\k|\d_{\k\beta}0_\k\>].\label{C1}
\end{equation}
where the symbol $\d_{\k,\alpha}$ means taking the derivative of $\alpha$ component of $\mathbf{B}_\k$ as defined by Eq.(\ref{partial derivative}). The total Berry curvature is defined as:
\begin{equation}
   \mathcal{F}_{\alpha\beta}=-2\mathbf{Im}[\<\d_\alpha 0|\d_\beta 0\>],
\end{equation}
where $|0\>$ means the total ground state of the total Hamiltonian which can be written as the direct product of $\k$-component ground state as:
\begin{equation}
   |0\>=\bigotimes_{\k=1}^{n}|0_{\k}\>=|0_{1}\>\otimes|0_{2}\>\otimes\cdots\otimes|0_{n}\>=|0_{1}\>|0_{2}\>\cdots|0_{n}\>
\end{equation}
Then we substitute the above  equation into Eq.(\ref{C1}) and we obtain
\begin{equation}
   \begin{split}
      \mathcal{F}_{\alpha\beta}&=-2\mathbf{Im}\left[ \sum_{\k,\q}\left(\d_{\k,\alpha}\<0_{1}|\cdots\<0_{n}|\right)\cdot(\d_{\q,\beta}|0_{1}\>\cdots|0_{n}\>) \right]\\
      &=-2\mathbf{Im}\left[
         \sum_\k\<\d_{\k,\alpha}0_{\k}|\d_{\k,\beta}0_\k\>+\sum_{\k}\sum_{\q\neq\k}\<\d_{\k,\alpha}0_\k|0_\k\>\cdot\<0_\q|\d_{\q,\beta}|0_\q\>
      \right]=\sum_{\k}\mathcal{F}_{\k,\alpha\beta}-2\mathbf{Im}\left[\sum_{\k}\sum_{\q\neq\k}\mathcal{A}^*_{\k\alpha}\mathcal{A}_{\q\beta}\right]
      .\label{C2}
   \end{split}
\end{equation}
with
$$ \mathcal{A}_{\k\alpha}=-i\<0_\k|\d_{\k,\alpha}|0_\k\>$$
Here, we'll try to demonstrate that the second term of the square bracket will vanish because it is real, instead the first term will be remained. For the second term,  we can use the normalization of the eigenstate:$\<0_\k|0_\k\>=1$ and we defferentiate it:
\begin{equation}
   \d_{\k,\alpha}\<0_\k|0_\k\>=\<\d_{\k,\alpha}0_\k|0\>+\<0_\k|\d_{\k,\alpha}0_\k\>=0\Rightarrow \<\d_{\k,\alpha}0_\k|0_\k\>=-\<0_\k|\d_{\k,\alpha}0_\k\>.\label{CC1}
\end{equation}
it can be noticed that$ \<\d_{\k,\alpha}0_\k|0_\k\>$ and $\<0_\k|\d_{\k,\alpha}0_\k\>$ are conjugate to each other then  we can claim that both of them are purely imaginary. Then the second term of above will vanish. As for the first term, we can insert a set of complete basis by:
\begin{equation}
   \<\d_{\k,\alpha} 0_\k|\d_{\k,\beta}0_\k\>
   =\<\d_{\k,\alpha}0_\k|0_\k\>\<0_\k|\d_{\k,\beta}0_\k\>+\<\d_{\k,\alpha}0_\k|1_\k\>\<1_\k|\d_{\k,\beta}0_\k\>\label{CC2}
\end{equation}
As we analysed in Eq.(\ref{CC1}), the first term of Eq.(\ref{CC2}) is real and we can ignore it.
Then we analyse the second term of Eq.(\ref{CC2}). With the orthonormal property of the eigenstate $\<0_\k|1_\k\>=0$ we obtain
\begin{equation}
   \d_{\k,\alpha}\<0_\k|1_\k\>=0=\<\d_{\k,\alpha}0_\k|1_\k\> +\<0_\k|\d_{\k,\alpha}1_\k\>\Rightarrow \<\d_{\k,\alpha}0_\k|1_\k\>=-\<0_\k|\d_{\k,\alpha}1_\k\>.
\end{equation}
But we notice that $\<\d_{\k,\alpha}0_\k|1_\k\>$ and $\<0_\k|\d_{\k,\alpha}1_\k\>$ are not conjugate to each other, so they are not purely imaginary and the first term of Eq.(\ref{C2}) will not vanish. Hence, we derive the total Berry curvature can be written as the summation of the spectral Berry curvature over $\k$ as:
\begin{equation}
   \mathcal{F}_{\alpha\beta}=\sum_\k \mathcal{F}_{\k,\alpha\beta}.
\end{equation}

\subsection*{C. Results of quantum geometric tensor}
We analytically calculated the quantum geometric tensor of the Kitaev spin liquid and the results are given as follows
\begin{equation}
   \begin{split}
      Q_{\k,xx}&=\frac{\sin^2\phi_\k+\cos^2\theta_\k\cos^2\phi_\k}{4\varepsilon_\k^2}=\frac{\frac{\left(\frac{\xi_\k}{\alpha_\k}\right)^2}{1+\left(\frac{\xi_\k}{\alpha_\k}\right)^2}+\frac{\xi_\k^2}{\alpha_\k^2+\beta_\k^2+\xi_\k^2}\frac{1}{1+\left(\frac{\xi_\k}{\alpha_\k}\right)^2}}{4(\alpha_\k^2+\beta_\k^2+\xi_\k^2)}\\
      Q_{\k,xy}&=\frac{-\sin\theta_\k\sin\phi_\k\sin\theta_\k\cos\phi_\k-i\cos\theta_\k}{4\varepsilon_\k^2}=\frac{\alpha_\k\beta_\k-i\xi_\k(\alpha_\k^2+\beta_\k^2+\xi_\k^2)^{\frac12}} {4(\alpha_\k^2+\beta_\k^2+\xi_\k^2)}\\
      Q_{\k,xz}&=\frac{-\sin\theta_\k\cos\phi_\k\cos\theta_\k+i\sin\theta_\k\sin\phi_\k}{4\varepsilon_\k^2}=\frac{\alpha_\k\xi_k-i\beta_\k(\alpha_\k^2+\beta_\k^2+\xi_\k^2)^{\frac12}}{(\alpha_\k^2+\beta_\k^2+\xi_\k^2)^2}\\
      Q_{\k,yy}&=\frac{\cos^2\theta_\k\sin^2\phi_\k+\cos^2\theta_\k}{4\varepsilon_\k^2}=\frac{\frac{1}{1+\left(\frac{\xi_\k}{\alpha_\k}\right)^2} \left(\frac{\beta_\k}{\alpha_\k}\right)^2+\frac{1}{1+\left(\frac{\xi_\k}{\alpha_\k}\right)^2}}{4(\alpha_\k^2+\beta_\k^2+\xi_\k^2)}\\
      Q_{\k,yz}&=\frac{-\sin\theta_\k\sin\phi_\k\cos\theta_\k-i\sin\theta_\k\cos\phi_\k}{4\varepsilon_\k^2}=\frac{\beta_\k\xi_\k-i\alpha_\k(\alpha_\k^2+\beta_\k^2+\xi_\k^2)^\frac12}{4(\alpha_\k^2+\beta_\k^2+\xi_\k^2)^2}\\
      Q_{\k,zz}&=\frac{\sin\theta_\k^2}{4\varepsilon_\k^2}=\frac{\frac{\left(\frac{\xi_\k}{\alpha_\k}\right)^2}{1+\left(\frac{\xi_\k}{\alpha_\k}\right)^2}}{4(\alpha_\k^2+\beta_\k^2+\xi_\k^2)}.
      \label{explicit form of tensor}
   \end{split}
\end{equation}

\subsection*{D. 2-order nonlinear magnetic susceptibility}
The second order magnetic susceptibility is defined as: (for brevity we omitted the index $\k$)
\begin{equation}
    \chi_{\alpha\beta\gamma}(t,t',0)=(-i)^2\Theta(t-t')\Theta(t')\<[[\sigma_{\alpha}(t),\sigma_\beta(t')],\sigma_\gamma(0)]\>_0
\end{equation}
For the sake of simplicity, we denote $\sigma_\alpha,\sigma_\beta,\sigma_\gamma $ by $A,B,C$ respectively. Then we insert two sets of complete bases inside the above equation. Because the commutator will be expanded for four terms, here we only present the first term
\begin{equation}
    \begin{split}
        \chi_{\alpha\beta\gamma}(t,t',0)=-\Theta(t-t')\Theta(t')(&\<0|A(t)|0\>\<0|B(t')|0\>\<0|C|0\>+\<0|A(t)|1\>\<1|B(t')|0\>\<0|C|0\>\\
        +&\<0|A(t)|0\>\<0|B(t')|1\>\<1|C|0\>+\<0|A(t)|1\>\<1|B(t')|1\>\<1|C|0\>
        )
    \end{split}
\end{equation}
Then we use $A_{mn}$ to denote the element of the matrix. Then the above can be written briefly
\begin{equation}
    \begin{split}
        \chi_{\alpha\beta\gamma}(t,t',0)&=-\Theta(t-t')\Theta(t')\times\\
        [
            &(A_{00}(t)B_{00}(t')C_{00}+A_{01}(t)B_{10}(t')C_{00}+A_{00}(t)B_{01}(t')C_{10}+A_{01}(t)B_{11}(t')C_{10})\\
            -&(B_{00}(t')A_{00}(t)C_{00}+B_{01}(t')A_{10}(t)C_{00}+B_{00}(t')A_{01}(t)C_{10}+B_{01}(t')A_{11}(t)C_{10})\\
            -&(C_{00}A_{00}(t)B_{00}(t')+C_{01}A_{10}(t)B_{00}(t')+C_{00}A_{01}(t)B_{10}(t')+C_{01}A_{11}(t)B_{10}(t') )\\
            +&(C_{00}B_{00}(t')A_{00}(t)+C_{01}B_{10}(t')A_{00}(t)+C_{00}B_{01}(t')A_{10}(t)+C_{01}B_{11}(t')A_{10}(t))
        ]
    \end{split}
\end{equation}
Under the Heisenberg picture, we find that
\begin{equation}
    A(t)=\mathbf{e}^{iHt}A\mathbf{e}^{-iHt}\Rightarrow A_{00}(t)=\<0|A|0\>=A_{00},
\end{equation}
and the same for $B(t')$ and we can cancle some terms. Then the second order magnetic susceptibility can be simplified as:
\begin{equation}
    \begin{split}
        \chi_{\alpha\beta\gamma}(t,t',0)&=-\Theta(t-t')\Theta(t')\times\\
        [
           &\mathbf{e}^{-i2\epsilon t'}(A_{00}B_{01}C_{10}-A_{11}B_{01}C_{10})+\mathbf{e}^{i2\epsilon t'}(A_{00}B_{10}C_{01}-A_{00}B_{10}C_{01})\\
           +&\mathbf{e}^{-2i\epsilon t}(A_{01}B_{11}C_{10}-A_{01}B_{00}C_{10})+\mathbf{e}^{i2\epsilon t}(A_{10}B_{11}C_{01}-A_{10}B_{00}C_{01})
        ]\\
        &=-\Theta(t-t')\Theta(t')(\mathbf{e}^{-i2\epsilon t'}f_1+\mathbf{e}^{i2\epsilon t'}f_2+\mathbf{e}^{-i2\epsilon t}f_3+\mathbf{e}^{i2\epsilon t}f_4)
    \end{split}
\end{equation}
where we consider two-level model with two eigenenergies $\pm\epsilon$, then we use the integral expression of $\Theta $  function as:
\begin{equation}
    \Theta(t)=-\frac{1}{2\pi i}\int_{-\infty}^\infty \frac{\mathbf{e}^{ixt}}{x+i\eta}~\mathrm{d}x.
\end{equation}
and use the Lehmann representation by means of Fourier transformation:
\begin{equation}
    \begin{split}
        \chi_{\alpha\beta\gamma}(\omega,\omega')&=-\left(\frac{i}{2\pi}\right)^2\int_{4-D}\mathrm{d}x\mathrm{d}y\mathrm{d}t\mathrm{d}t'\left[\frac{\mathbf{e}^{ix(t-t')}\mathbf{e}^{iyt'}\mathbf{e}^{i\omega t}\mathbf{e}^{i\omega't'}}{(x+i\eta)(y+i\eta')}(\mathbf{e}^{-i2\epsilon t'}f_1+\mathbf{e}^{i2\epsilon t'}f_2+\mathbf{e}^{-i2\epsilon t}f_3+\mathbf{e}^{i2\epsilon t}f_4)\right]
    \end{split}
\end{equation}
Then we use the integral expression of $\delta$ function as
\begin{equation}
    \delta(u-u_0)=\frac{1}{2\pi}\int_{-\infty}^\infty\mathrm{d}t~ \mathbf{e}^{it(u-u_0)}
\end{equation}
and we'll obtain the brief expression of $\chi_{\alpha\beta\gamma}(\omega)$:
\begin{equation}
    \begin{split}
        &\chi_{\alpha\beta\gamma}(\omega,\omega')\\
        &=\iint\mathrm{d}x\mathrm{d}y\left[\frac{1}{(x+i\eta)(y+i\eta')}\right] [\delta(-x+y+\omega'-2\epsilon)\delta(x+\omega)f_1\\
        &+\delta(-x+y+\omega'+2\epsilon)\delta(x+\omega)f_2+\delta(x+\omega-2\epsilon)\delta(-x+y+\omega')f_3+\delta(x+\omega+2\epsilon)\delta(-x+y+\omega')f_4]\\\\
        &=\frac{\<0|A|0\>\<0|B|1\>\<1|C|0\>-\<1|A|1\>\<0|B|1\>\<1|C|0\>}{(-\omega+i\eta)(-\omega-\omega'+2\epsilon+i\eta')}+\frac{\<0|A|0\>\<1|B|0\>\<0|C|1\>-\<1|A|1\>\<1|B|0\>\<0|C|1\>}{(-\omega+i\eta)(-\omega-\omega'-2\epsilon+i\eta')}\\
        &+\frac{\<0|A|1\>\<1|B|1\>\<1|C|0\>-\<0|A|1\>\<0|B|0\>\<1|C|0\>}{(-\omega+2\epsilon+i\eta)(\omega'-\omega+2\epsilon+i\eta')}+\frac{\<1|A|0\>\<1|B|1\>\<0|C|1\>-\<1|A|0\>\<0|B|0\>\<0|C|1\>}{(-\omega-2\epsilon+i\eta)(\omega'-\omega-2\epsilon+i\eta')}
    \end{split}
\end{equation}
Under the natural  unit, the frequency has the same dimension as the energy, so we let $\omega',\omega$ take such values as $n\epsilon_\k,m\epsilon_\k$, respectively and the susceptibility can be written as:
\begin{equation}
   \begin{split}
      &\chi_{\k,\alpha\beta\gamma}(\omega,\omega')\\
      &=\frac{\<0|A|0\>\<0|B|1\>\<1|C|0\>-\<1|A|1\>\<0|B|1\>\<1|C|0\>}{m(m+n-2)\epsilon_\k^2}+\frac{\<0|A|0\>\<1|B|0\>\<0|C|1\>-\<1|A|1\>\<1|B|0\>\<0|C|1\>}{m(m+n+2)\epsilon_\k^2}\\
      &+\frac{\<0|A|1\>\<1|B|1\>\<1|C|0\>-\<0|A|1\>\<0|B|0\>\<1|C|0\>}{(-m+2)(n-m+2)\epsilon_\k^2}+\frac{\<1|A|0\>\<1|B|1\>\<0|C|1\>-\<1|A|0\>\<0|B|0\>\<0|C|1\>}{(-m-2)(n-m-2)\epsilon_\k^2}\\
      &=\frac{4}{m(m+n-2)}(\<\sigma_{\k,\alpha}\>_{\k,0}\mathcal{A}_{\k,\beta}^{01}\mathcal{A}_{\k,\gamma}^{10}-\<\sigma_{\k,\alpha}\>_{\k,1}\mathcal{A}_{\k,\beta}^{01}\mathcal{A}_{\k,\gamma}^{10})\\
      &+\frac{4}{m(m+n+2)}(\<\sigma_{\k,\alpha}\>_{\k,0}\mathcal{A}_{\k,\beta}^{10}\mathcal{A}_{\k,\gamma} ^{01}-\<\sigma_{\k,\alpha}\>_{\k,1}\mathcal{A}_{\k,\beta}^{10}\mathcal{A}_{\k,\gamma}^{01})\\
      &+\frac{4}{(m-2)(n-m+2)}(\<\sigma_{\k,\beta}\>_{\k,1}\mathcal{A}_{\k,\alpha}^{01}\mathcal{A}_{\k,\gamma}^{10}-\<\sigma_{\k,\beta}\>_{\k,0}\mathcal{A}_{\k,\alpha}^{01}\mathcal{A}_{\k,\gamma}^{10})\\
      &+\frac{4}{(m+2)(n-m-2)}(\<\sigma_{\k,\beta}\>_{\k,1}\mathcal{A}_{\k,\alpha}^{10}\mathcal{A}_{\k,\gamma}^{01}-\<\sigma_{\k,\beta}\>_{\k,0}\mathcal{A}_{\k,\alpha}^{10}\mathcal{A}_{\k,\gamma}^{01})\label{second order}
  \end{split}
\end{equation}
where $m, n$ are real and they're defined as:
\[
m=\frac{\omega}{\epsilon_\k},~~~n=\frac{\omega'}{\epsilon_\k}.
\]


\end{widetext}

\end{document}